\documentclass[twocolumn,showpacs,preprintnumbers,amsmath,amssymb]{revtex4}

\usepackage{graphicx}
\usepackage{dcolumn}
\usepackage{bm}

\newcommand{\nc}{\newcommand}

\newcommand{\he}{\hat{e}}
\newcommand{\hv}{\hat{v}}
\newcommand{\vs}{\vec{\sigma}}

\nc{\ra}{\rightarrow}

\nc{\da}{\dagger}
\nc{\dn}{\downarrow}
\nc{\cA}{{\cal A}}
\nc{\cB}{{\cal B}}
\nc{\cC}{{\cal C}}
\nc{\cD}{{\cal D}}
\nc{\cE}{{\cal E}}
\nc{\cF}{{\cal F}}
\nc{\cG}{{\cal G}}
\nc{\cH}{{\cal H}}
\nc{\cI}{{\cal I}}
\nc{\cJ}{{\cal J}}
\nc{\cK}{{\cal K}}
\nc{\cL}{{\cal L}}
\nc{\cR}{{\cal R}}
\nc{\cS}{{\cal S}}
\nc{\cX}{{\cal X}}
\nc{\cO}{{\cal O}}
\nc{\cU}{{\cal U}}

\begin{document}


\title{Unsolvability of the Halting Problem in Quantum Dynamics}

\author{Daegene Song}
\affiliation{National Institute of Standards and Technology, Gaithersburg, Maryland 20899}%

\date{\today}

\begin{abstract}
It is shown that the 
halting problem cannot be solved consistently in both 
 the Schr\"odinger and Heisenberg pictures of 
quantum dynamics.  
 The existence of the halting machine, 
 which is assumed from quantum theory, 
 leads into a contradiction  
when we consider the case when the 
observer's reference frame is the system 
that is to be evolved in 
both pictures. 
We then show that in order to include the evolution of 
observer's reference frame in a physically sensible way, 
the Heisenberg picture with time
going backwards yields a correct description. 

\end{abstract}

\maketitle
With the construction of universal quantum Turing machine, 
Deutsch proposed \cite{deutsch} a quantum version of the halting problem 
first proved by Alan Turing in 1936 \cite{turing}.
In recent years, a lot of interest has been focused on quantum computation \cite{chuang},   
and the discussion of the halting problem using 
a quantum computer has also received attention. 
Myers argued \cite{myers} that due to entanglement 
between a halt qubit and a system, it may be difficult to 
measure the halt qubit, which may spoil the computation.  
Subsequent discussions on the halting problem 
with a quantum computer have mainly focused 
on the superposition and entanglement of the halt qubit \cite{ozawa,linden,shi}.  
In this paper, we approach the halting scheme differently 
and use two pictures of 
quantum dynamics, i.e., the Schr\"odinger and Heisenberg pictures.  
Schr\"odinger's wave mechanics and Heisenberg's matrix mechanics 
were formulated 
in the early twentieth century and have been considered to be equivalent, 
i.e., two different ways of describing the same physical phenomenon that we observe.  
Therefore, in order to consider a halting scheme 
for a quantum system, we need to 
examine whether the procedure is consistent 
in both the Schr\"odinger and Heisenberg pictures. 
We will give an example in quantum dynamics that 
shows this cannot be achieved. 
We will then argue that it is the Heisenberg picture, 
rather than both pictures, that yields the 
correct description that not only does not run into the inconsistency 
shown through the halting scheme but also is physically sensible.

\begin{figure}
\begin{center}
{\includegraphics[scale=.4]{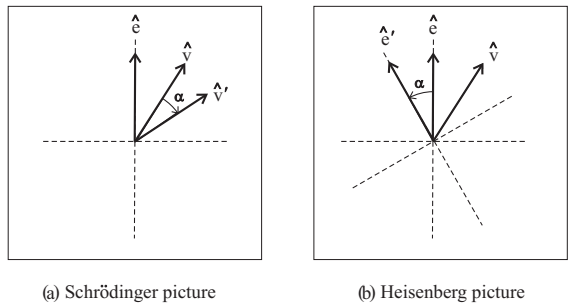}}
\end{center}
\caption{For the Schr\"odinger picture (a), the unit vector ${\bf {\hv}}$ 
evolves while the unit basis vector ${\bf {\he}}$ is intact.  
In the Heisenberg picture (b), the basis vector ${\bf {\he}}$ 
is rotated into opposite direction by the same amount 
while the unit vector  ${\bf {\hv}}$ remains, thereby keeping the 
angle between the two vectors, therefore the expectation values, the same in both pictures. }
\label{heisenberg}\end{figure}

In order to discuss the halting problem, we first wish to define notations to be used.   
In particular we will follow a similar notation used in 
 \cite{gottesman,hayden} such that it is convenient in both Schr\"odinger and Heisenberg pictures.  
A qubit, a basic unit of quantum information, is a two-level quantum system written as   
$|\psi\rangle =a|0\rangle + b|1\rangle$.  
Using a Bloch sphere notation, i.e., with $a =\exp(-i\phi/2)\cos(\theta/2)$ 
and $b =\exp(i\phi/2)\sin(\theta/2)$, a qubit in a density matrix form can be written as 
$|\psi\rangle\langle \psi| = \frac{1}{2}({\bf {1}}+ {\bf{\hv}} \cdot \vs )$ where 
 $({\bf {v}}_x,{\bf {v}}_y,{\bf {v}}_z)$ 
 $=(\sin\theta \cos\phi,\sin\theta\sin\phi,\cos\theta )$ and  
$\vs = (\sigma_x,\sigma_y,\sigma_z)$ with 
$\sigma_x = |0\rangle\langle 1| + |1\rangle\langle 0|$, 
$\sigma_y = -i|0\rangle\langle 1| + i|1\rangle\langle 0|$, 
and $\sigma_z = |0\rangle\langle 0| - |1\rangle\langle 1|$.   
Therefore a qubit, $|\psi\rangle\langle \psi|$,    
can be represented as a unit vector   
${\bf{\hv}} = ({\bf {v}}_x,{\bf {v}}_y,{\bf {v}}_z)$
pointing in $(\theta,\phi)$ of a sphere with $0\leq \theta \leq \pi , 0\leq \phi \leq 2\pi$. 
A unitary transformation of a qubit in the unit vector notation ${\bf{\hv}}$ can be obtained by applying 
$U$ to $\sigma_i$ for the corresponding $i$th component of the vector 
${\bf{\hv}}$, i.e., ${\bf {v}}_i$, where $i=x,y,z$ 
(also see \cite{hardy} for a general transformation of a single qubit in a Bloch sphere). 
We will write the transformation of ${\bf {\hv}}$ under the 
unitary operation $U$ as ${\bf {\hv}}^{\prime} = U {\bf {\hv}}U^{\dagger}$,
 implying the unitary transformation is applied to the corresponding $\sigma_i$. 
For example, let us consider the case when $U$ is a rotation about $y$-axis by $\alpha$ in a Bloch sphere, i.e., 
$U = \cos\frac{\alpha}{2}|0\rangle\langle 0| -\sin\frac{\alpha}{2}|0\rangle\langle 1| +\sin\frac{\alpha}{2} |1\rangle\langle 0| + \cos\frac{\alpha}{2}|1\rangle\langle 1|$.
Then it yields that ${\bf {\hv}} = ({\bf{v}}_x,{\bf{v}}_y,{\bf{v}}_z)$ is transformed into 
${\bf {\hv}}^{\prime} $  $\equiv  U {\bf {\hv}}U^{\dagger} $ 
   = $  \left( \cos \alpha {\bf {v}}_x + \sin \alpha {\bf {v}}_z,{\bf{v}}_y, -\sin\alpha {\bf{v}}_x + \cos \alpha {\bf{v}}_z \right)$.  
In quantum theory, there is another important variable called an observable.  
For a single qubit, an observable can also be written as a unit vector \cite{hayden},     
${\bf{\he}}=( {\bf{e}}_x,{\bf {e}}_y,{\bf {e}}_z )$
where $({\bf {e}}_x,{\bf {e}}_y,{\bf {e}}_z)$ 
$=(\sin\vartheta \cos\varphi,\sin\vartheta\sin\varphi,\cos\vartheta )$, 
pointing $(\vartheta,\varphi)$ direction in a sphere.  
Therefore if one is to make a measurement in $(\vartheta,\varphi)$ direction, 
 the observable would be ${\bf {\he}} \cdot \vs$. 
In the Heisenberg picture of quantum theory, it is the unit basis vector
${\bf{\he}}$ that is transformed (p243, \cite{peres}). 
Using a similar transformation rule as in ${\bf {\hv}}$, 
a unitary transformation of the observable in the basis vector notation can be obtained by applying 
$U^{\dagger}$ to the $\sigma_j$ by $U^{\dagger} \sigma_j U$ for ${\bf {e}}_j$ which we represent 
 as ${\bf {\he}}^{\prime} = U^{\dagger} {\bf {\he}} U$. 
As an example, we again consider the case when $U$ is a rotation about $y$-axis by $\alpha$ as follows 
${\bf {\he}}^{\prime}$    $ \equiv  U^{\dagger} {\bf {\he}} U$ 
  = $ \left( \cos \alpha {\bf {e}}_x -\sin\alpha {\bf {e}}_z,{\bf {e}_y}, \sin\alpha {\bf {e}}_x + \cos\alpha{\bf {e}}_z \right)$.  
As shown in Fig. \ref{heisenberg}, the directions of transformation 
for two vectors are different for Schr\"odinger and Heisenberg pictures.  
Therefore the expectation value ${\bf {\he}}^{\prime}\cdot {\bf {\hv}}$ in the Heisenberg picture 
remains the same as in the case with the Schr\"odinger picture, i.e., $ {\bf {e}}\cdot{\bf {\hv}}^{\prime} $.
For the remainder of this paper, 
we will treat the two vectors ${\bf{\hv}}$ and ${\bf{\he}}$ 
on an equal footing.  
The only specialty about ${\bf{\he}}$ is that 
it serves as a coordinate or a basis vector such that 
when a measurement is made on the vector ${\bf {\hv}}$, 
the expectation value is with respect to 
${\bf{\he}}$.  

With a quantum system and a halt qubit, 
Deutsch introduced \cite{deutsch} a quantum version 
of the halting problem wherein   
the completion of every valid quantum algorithm 
through a unitary process applied to 
the quantum system is accompanied by the 
change in a halt qubit to 1 that remains 0 otherwise. 
We will assume such a halting machine exists and will 
 argue that this assumption leads into a contradiction. 
With the introduced notations, 
we will consider one particular case of the halting machine,   
that is, when the halting machine consists of 
a unit vector ${\bf{\hv_s}} \equiv({\bf{v}}_x,{\bf{v}}_y,{\bf{v}}_z)$ and 
a halt qubit ${\bf{\hv_h}} \equiv (0,0,1)$.  
We do not include an ancilla state 
because it will not be needed for our discussion.  
The time evolution of the halting machine is defined through 
a unitary process, and the machine halts 
when the unit vector ${\bf{\hv_s}}$ is rotated   
 by $\delta$ about 
an arbitrary ${\hat{n}}=(n_x,n_y,n_z)$-axis.  
This time evolution of the halting machine can be 
achieved with the unit vector ${\bf{\hv_s}}$ evolving as follows 
\begin{equation} 
{\bf {\hv_s}} \rightarrow U_{\delta}{\bf {\hv_s}} U_{\delta}^{\dagger}
\label{HM}\end{equation}
 where $U_{\delta}\equiv \cos(\delta/2){\bf {1}}-i\sin(\delta/2)({\hat{n}}\cdot {\vec{\sigma}}) $
and the halt qubit ${\bf{\hv_h}}$ is transformed into 
$-{\bf{\hv_h}}$ with a unitary operation $\sigma_x$.  
 In the following, we will show  
 this halting machine  
runs into a contradiction.

Before we proceed with our discussion of the halting problem,  
we wish to discuss the concept of observables in quantum theory.  
 When we want to check
a moving vehicle's speed, we may use 
a speed gun and could read, for example,
80km/hr.  Or we could use a thermometer to 
measure a room temperature which may 
yield, for example, 25 degrees Celsius.
 While the measurement tools, 
such as the speed gun and the thermometer, yield the output with not
only numbers but also units such as km/hr and degrees Celsius, 
what the actual measurement yields is rather different. 
 For example, a laser speed gun checks the distances from the gun 
at two different times and is designed to 
calculate and to yield an output of 
the moving vehicle's speed.  
A mercury-thermometer is designed to show 
the temperature in relation to the increase of the 
volume of mercury in the thermometer.  
The numbers obtained from the measurement represent 
the perception experienced by an observer and 
the meaning of those numbers, such as speed or 
temperature represented with units, a concept, 
is imposed by an observer. 
In quantum theory, concepts such as position and 
momentum are called observables 
and the numbers that result from the measurements are 
represented as eigenvalues (p63, \cite{peres}).

\begin{figure}
\begin{center}
{\includegraphics[scale=.9]{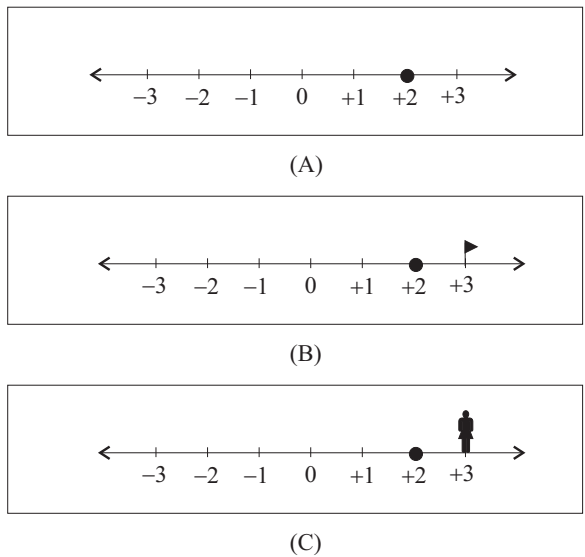}}
\end{center}
\caption{For (A), it is not possible to claim the black dot is on the right 
or on the left.  In (B), we may say, with respect to the flag in $+3$, 
the dot is on the left.  If we assume there is an observer 
living and sitting at $+3$ while facing into the paper 
(i.e., the same direction as the reader of this paper) 
as in (C) and if the observer  
measures and obtains 
the result that the dot is on the left, 
then it is the observer who is serving the role of the flag in (B), 
i.e., as a reference point. }
\label{persons}\end{figure}

Let us take an example of a one-dimensional 
line as shown in Fig. \ref{persons}.  
In order to claim a dot, which is lying on the line,  
is either on the right or on the left, there should be 
a reference point.  For example, with respect to 
the origin or with respect to +3, 
one may say the dot is on the left 
or on the right.  
Instead of looking at the line from outside, 
suppose there is an observer being confined to the 
one-dimensional line facing into the paper as 
shown in (C) of Fig. \ref{persons}.  
The observer measures or perceives whether 
the dot is on the right or on the left.  
Depending on where the observer is sitting, the outcome of the measurement,
 i.e., either on the right or on the left, 
will change.  
In this case, we note that 
the observer him or herself is serving the role of the reference point. 
Therefore when the observer makes a 
measurement and gets a result that the dot is on the 
right or on the left, this implies that with respect to 
his or her reference frame of the position on the line, 
the dot is on the right or on the left.  
Let us apply the same logic to the case 
of a single qubit in a Bloch sphere.  
When an observer measures 
a qubit in a certain direction, say in ${\hat{n}}$, 
the outcome of the measurement is either 
$+1$ or $-1$.  
The eigenvalue obtained is with respect to 
the measurement direction ${\hat{n}}$.  
It is noted that ${\hat{n}}$ is playing 
a similar role as the reference point 
 in the case of the one-dimensional line example.   
We also note that the measurement outcome of $+1$ or $-1$ 
is the perception experienced by the observer.   
That is, it is the observer who obtains the outcome $+1$ or $-1$. 
Therefore, the outcome should be meaningful 
with respect to the observer's certain reference frame. 
Because we already know that the eigenvalue 
outcome $+1$ or $-1$ is meaningful 
with respect to the measurement direction ${\hat{n}}$, 
it leads us to consider 
the observer's reference frame as ${\hat{n}}$ 
for our single qubit measurement case.   
Using the unit vector notations we previously defined, 
we propose the following: 
\\
{\bf{Postulate I}}: {\it{Given a unit vector ${\bf{\hv}}$, 
an observer's reference frame is identified with  
a basis unit vector ${\bf{\he}}$}}. 
\\
With this postulate, 
two pictures of quantum theory can have a 
natural physical realization between 
an observer and a system. 
Fig. \ref{heisenberg} shows that, in the Schr\"odinger picture, 
the observer's 
reference frame, represented by the unit basis vector 
${\bf {\he}}$, stays still while the state vector is   
rotated clockwise by $\alpha$, and the Heisenberg picture 
shows the unit vector 
stays still and the observer's coordinate is rotated 
counterclockwise by $\alpha$. 
In both cases, the observer would observe exactly the same phenomenon.

It should be noted that we are not using a notion 
of detector or apparatus in the place of an observer.  
According to our postulate, for a given unit vector, 
the observer's reference frame is represented with a 
unit basis vector in a Bloch sphere. 
However, it was shown that \cite{nielsen} a finite 
dimensional detector cannot 
encode an arbitrary unitary transformation whereas,  
according to our postulate, the observer's 
identified coordinate unit basis vector represents an 
arbitrary measurement basis for a given qubit.  
Therefore, we do not use the term detector or 
an apparatus to replace an observer.   
If one wants to include an apparatus or detector, we may consider 
the state, i.e., ${\bf {\hv}}$, to be a larger system 
that includes a qubit and an 
apparatus and the coordinate vector for an observer would also 
be represented by the same larger basis vector.  However, in this paper, 
we only consider the simplest possible case of a single qubit.

Let us now consider a system 
with an observer and the halting machine defined with the evolution in (\ref{HM}).  
That is, we are considering a closed system consisted of 
a quantum state, represented by the unit vector ${\bf{\hv_s}}=({\bf {v}}_x,{\bf {v}}_y,{\bf {v}}_z)$, 
an observer, whom we call Alice, represented by the reference frame 
${\bf{\he_s}} =({\bf {e}}_x,{\bf {e}}_y,{\bf {e}}_z)$  introduced above, 
and a halt qubit ${\bf{\hv_h}} $ along with Alice's reference frame for the halt qubit defined as 
${\bf{\he_h}} \equiv (0,0,1)$.  
Alice is to transform the unit vector ${\bf{\hv_s}}$ by $\delta$ about 
an arbitrary ${\hat{n}}=(n_x,n_y,n_z)$-axis with 
$U_{\delta}=\cos(\delta/2){\bf {1}}-i\sin(\delta/2)({\hat{n}}\cdot {\vec{\sigma}})$
 and also applies 
 $\sigma_x$ on a halt qubit such that $ {\bf{\hv_h}}  \rightarrow -{\bf{\hv_h}} $. 
If Alice were to measure the evolved vector state, the expectation value would be 
${\bf {\he_s}} \cdot (U_{\delta}{\bf {\hv_s}} U_{\delta}^{\dagger})$. 
Next, we wish to consider the same procedure in the Heisenberg picture. 
In the Schr\"odinger picture we discussed above, the unitary evolution was performed on ${\bf {\hv_s}}$. 
Therefore, in the Heisenberg picture, the $U_{\delta}^{\dagger}$ 
transforms the basis vector ${\bf{\he_s}}$ into 
$U_{\delta}^{\dagger}{\bf{\he_s}} U_{\delta}$ where 
$U_{\delta}^{\dagger}=\cos(\delta/2){\bf{1}}+i\sin(\delta/2)({\hat{n}}\cdot {\vec{\sigma}})$ 
and the observable for the halt qubit, i.e., ${\bf{\he_h}}$, is transformed into $-{\bf{\he_h}}$. 
It yields the expectation value of 
$(U_{\delta}^{\dagger} {\bf{\he_s}} U_{\delta})\cdot {\bf{\hv_s}}$ which is 
equal to the expectation value in the Schr\"odinger picture, 
${\bf{\he_s}} \cdot (U_{\delta} {\bf{\hv_s}} U_{\delta}^{\dagger})$.

We now consider the halting machine in (\ref{HM}) with one particular input.    
That is, when the input state to be transformed is the Alice's unit basis vector, i.e., ${\bf{\hv_s}}={\bf{\he_s}}$.  
Note that we are treating ${\bf{\hv}}$ and ${\bf{\he}}$ on a equal footing.   
  In the Schr\"odinger picture, the evolution is, 
${\bf{\he_s}} \rightarrow $  $U_{\delta} {\bf{\he_s}} U_{\delta}^{\dagger} \equiv {\bf{\he_s}}^{\prime\prime}$, 
and Alice also transforms ${\bf{\hv_h}} \rightarrow -{\bf{\hv_h}} $. 
We now consider the same procedure in the Heisenberg picture.  
In this case, the unit basis vector ${\bf{\he_s}}$, is transformed as   
${\bf{\he_s}} \rightarrow $  $U_{\delta}^{\dagger} {\bf{\he_s}} U_{\delta} \equiv {\bf{\he_s}}^{\prime\prime\prime}$ 
and $ {\bf{\he_h}} \rightarrow -{\bf{\he_h}} $.  
Note that ${\bf{\he_s}}^{\prime\prime} \neq {\bf{\he_s}}^{\prime\prime\prime}$ 
unless ${\bf{\he_s}} = \pm {\hat{n}}$ or $\delta= k\pi$ where $k=0,1,2...$. 
For the example of a system with an observer and the halting machine studied in the previous paragraph, 
the vector ${\bf{\hv_s}}$ has evolved, with respect to ${\bf{\he_s}}$, 
into the same output in both Schr\"odinger and 
Heisenberg pictures. Similarly, with respect to ${\bf{\he_h}}$, the halt qubit, ${\bf{\hv_h}}$, 
halted in both pictures.  
However, in the case with ${\bf{\he_s}}$ as an input we just considered, 
while the halt qubit ${\bf{\hv_h}}$ halted  
on both occasions with respect to ${\bf{\he_h}}$, the vector that is being evolved, i.e., ${\bf{\he_s}}$, turned out 
as two generally different outputs in two pictures.  
This contradicts our assumption about the halting machine in (\ref{HM}) because the machine should yield 
an output that is a rotation of the input by $\delta$ about a $\hat{n}$-axis and is unique.

\begin{figure}
\begin{center}
{\includegraphics[scale=.4 ]{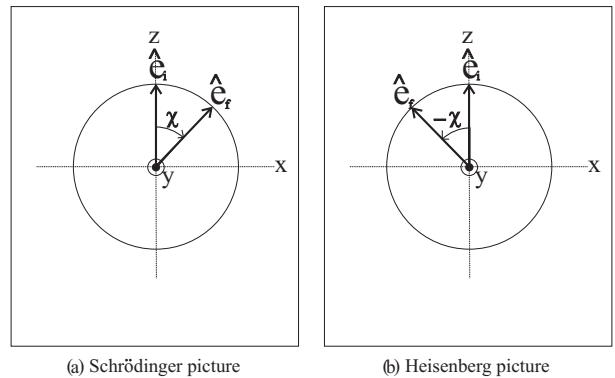}}
\end{center}
\caption{Unitary evolution of ${\bf{\he}}$ is considered. The vector ${\bf{\he}}$ is initially pointing $z$-direction and 
is rotated about $y$-axis by $\chi$ after time $t$.  In the Schr\"odinger picture as in (a), the vector is rotated clockwise 
and in (b) the vector is rotated by $-\chi$.    }
\label{Sch}\end{figure}

Therefore, we have shown that the existence of the halting machine 
that is assumed from quantum theory leads into a contradiction 
when we consider the input of the unit basis vector 
${\bf{\he_s}}$ (for simplicity, we will omit the subscript $s$ from now on), 
which is transformed 
into two generally different outputs
in  Schr\"odinger and Heisenberg pictures.
However, not only can the halting problem not be solved consistently in 
both pictures, but also the evolution of the unit basis vector ${\bf{\he}}$ 
is physically sensible in neither of the two pictures in quantum dynamics. 
With our first postulate, 
we were able to impose a physical meaning on the 
Schr\"odinger and Heisenberg pictures of quantum theory.   
That is, in case of the Schr\"odinger picture, 
the system is evolving while an observer's reference frame is intact and,    
for the Heisenberg's picture, an observer's coordinate 
is evolving and the system is staying still.  
The equivalence of these two pictures comes from the fact that 
the observer would observe the same phenomenon 
and would not be able to tell the difference between them.  
For example, an observer applying a unitary operation 
to a qubit is experiencing a unitary evolution being applied to the qubit and this experience is the same 
in both pictures.  
But when it is the observer's reference frame that is evolving, 
it is difficult to imagine how an observer could observe or experience it.   
As shown in (a) of Fig. \ref{Sch}, let us assume that initially  
vector ${\bf{\he}}$ is pointing $z$-direction and with the 
unitary operation of rotation about $y$-axis, ${\bf{\he}}$ 
evolves under 
\begin{equation}
U=e^{-i \sigma_y t /2}
\end{equation}
in the Schr\"odinger picture.  
And the final state of ${\bf{\he}}$ 
would be rotated by $\chi$ after time $t$, 
which we write as $\chi(t)$.  
The difficulty with this evolution is that 
in order to experience the unitary evolution, 
Alice needs to be in another reference frame, say $\chi^{\prime}(t)$.   
However, ${\bf{\he}}$ itself is Alice's reference frame and 
there cannot be another reference frame.  
Similarly, in the Heisenberg picture, ${\bf{\he}}$ evolves under 
\begin{equation}
U^{\dagger}=e^{i \sigma_y t /2}
\label{heisenberg1}\end{equation}
As shown in Fig. \ref{Sch}, 
the vector is being rotated counterclockwise and is in 
$-\chi(t)$. In this case, 
for the observer in the reference frame of $-\chi(t)$, 
there needs to be additional vector 
in $\chi^{\prime}(t)$ in order for Alice to experience the evolution of 
${\bf{\he}}$.  
Again, this is not possible because $-\chi(t)$ 
is not only Alice's reference frame 
but also the system vector.    
Therefore, in order to have a satisfactory picture 
of Alice observing her own reference frame's evolution, 
Alice needs another reference frame  or another vector.

Therefore, it is not possible for either picture 
to be the correct way  
to describe the observer's experience of the evolution of ${\bf{\he}}$. 
Because ${\bf{\he}}$ is serving the 
role of both what the observer experiences 
and the observer's own reference frame, we need a picture such that 
the evolution of ${\bf{\he}}$ is neither of 
them yet yields the observer's experience of ${\bf{\he}}$'s evolution.
In order to resolve the dilemma discussed above and 
to determine the correct description for the observer's experience, 
we introduce our second postulate as follows:  
\newline
{\bf{Postulate II}}: {\it{What an observer observes or experiences 
must be time forwarding.}} 
\newline
Note that we are only postulating that 
the observer's experience is time forwarding 
and not necessarily the whole system, 
i.e., including the physical system and the observer, is time forwarding.

Let us re-consider the evolution of ${\bf{\he}}$ 
under the Heisenberg picture. 
Note that for the unitary operation in (\ref{heisenberg1}), 
it is possible to change 
the signs of $t$ and $\sigma_y$ while keeping 
the whole unitary operator the same, that is 
\begin{equation}
U^{\dagger}= e^{-i \sigma_y (-t)/2}
\label{minus}\end{equation}
This corresponds to the vector evolving to $\chi$ 
while $t$ is going to the minus direction compared to 
the previous Heisenberg case wherein the vector 
evolved to $-\chi$  with time going forward.  
In this case, we note that the observer cannot 
be in the reference frame $\chi(-t)$ because  
from the second postulate, 
we assumed what the observer observes or experiences is only time forwarding.  
If Alice is in the reference frame that is moving backward in time, 
she would observe everything going backward in time.  
However, from the assumption we made with the second postulate, 
this is not possible.  
We may consider the same trick with Schr\"odinger picture evolution, 
that is, by putting minus signs for both time 
and $\sigma_y$.  
But in this case, it still requires an additional 
observer's reference frame because the observer who is in the 
reference frame with time forwarding 
would simply observe $\chi$ in $+t$.  This is similar to the way 
an electron in the negative energy would appear as a positron 
in the positive energy to an observer who is also in the positive energy. 
Therefore, in the Schr\"odinger picture, 
this new view still requires an additional reference frame 
and is not satisfactory.

Therefore, with two postulates, 
in order to have a satisfactory description of experiencing 
the evolution of ${\bf{\he}}$ as well as of ${\bf{\hv}}$,  
we are forced to conclude that the quantum evolution follows 
according to the Heisenberg picture, not the Schr\"odinger picture,  
with time going backwards as shown in (\ref{minus}).    
Moreover, it leads us to abandon the general 
picture having the observer being 
in a certain reference frame evolving in time 
and observing the other vector.  
In other words, the more familiar picture of 
the observer being in the reference frame that is evolving forward in time 
should be abandoned, and the observer should be 
identified as what is being observed, i.e., $\chi$, 
 and its association with time, $t$.

\end{document}